# Oil drop deposition on solid surfaces in mixed polymer-surfactant solutions in relation to hair- and skin-care applications

Rumyana D. Stanimirova [a], Peter A. Kralchevsky [a,*], Krassimir D. Danov [a],
Hui Xu [b], Yee Wei Ung [b], Jordan T. Petkov [b,†]

[a] *Department of Chemical & Pharmaceutical Engineering, Faculty of Chemistry & Pharmacy, Sofia University, 1164 Sofia, Bulgaria.*

[b] *KL-Kepong Oleomas SDN BHD, Menara KLK, Jalan PJU 7/6, Mutiara Damansara, 47810 Petaling Jaya, Selangor Dalur Ehsan, Malaysia*

**Abstract**

The deposition of oil drops on solid substrates from mixed solutions of surfactants and cationic polymer is investigated. The used anionic surfactants are sodium laurylethersulfate (SLES) and sulfonated methyl esters (SME); the zwitterionic surfactant is cocamidopropyl betaine (CAPB). A new method, called the pressed drop method (PDM), was proposed to study the drop adhesion to substrates of different hydrophobicity. The PDM allows one to detect the presence or absence of drop adhesion at different degrees of dilution of the initial solution and, thus, to determine the threshold concentration of drop adhesion. The results show that the increase of the fraction of CAPB in the mixture with the anionic surfactant suppresses the oil-drop deposition; SME provides easier drop adhesion than SLES; the addition of NaCl enhances, whereas coco-fatty-acid-monoethanolamide (CMEA) suppresses the drop deposition; no drop adhesion is observed in the absence of polymer. The drop-to-substrate adhesion is interpreted in terms of the acting surface forces: polymer bridging attraction; hydrophobic attraction between segments of adsorbed polymer brushes and electrostatic forces. From viewpoint of applications, the PDM experiments enable one to compare the performance of various components in personal care formulations and to optimize their composition with respect to the oil-drop deposition.

Keywords:
Oil-drop deposition; Sulfonated methyl esters; Cationic polymer; Silicon oil; Shampoo formulations.

______________________________________________________________________

\* Corresponding author. Fax: +359 2 962 5643.
 *Email address*: pk@lcpe.uni-sofia.bg (P.A. Kralchevsky)

† Present address: Arch UK Biocides Ltd., Hexagon Tower, Crumpsall Vale, Blackley, Manchester M9 8GQ, UK

# 1. Introduction

Formulations used in hair- and skin-care products contain small oil drops, which improve softness and condition hair and skin [1-7]. Because shampoos and other personal care products must possess cleansing functions, they contain also anionic surfactants, which adsorb and bring negative surface charge to the oil drops and substrate. The resulting electrostatic repulsion suppresses the oil-drop deposition on the substrate [8]. To overcome this undesired effect, the respective personal-care formulations contain also a cationic polymer, which serves as a mediator of the drop-to-substrate adhesion [9-15]. In the bulk of solution, the surfactant and polymer form joint aggregates, sometimes called "coacervates" [16]. Both surfactants and polymers may adsorb on the surfaces of the oil drops and on the solid substrate. They are present also in the wetting films intervening between the oil drop and the substrate. For this reason, the surfactant–polymer interactions in the bulk and in the thin liquid films [15,17–21], are of primary importance for the oil-drop deposition. Electrostatic, hydrophobic, and polymer-bridging surface forces are expected to govern the oil-drop deposition [14,22,23].

The deposition process is illustrated in Fig. 1. The surfactant and polymer adsorb on the surface of the oil drop. Because the surfactant hydrophilizes the polymer and the drop, no oil deposition is observed at the higher concentrations. Upon dilution (rinsing), most of the surfactant is washed away, but the bigger and relatively more hydrophobic polymer molecules remain adsorbed on the oil drop and mediate its adhesion to the substrate.

The presence of polymer is crucial – as a rule no oil deposition is observed in the absence of polymer [8]. The polymer should adsorb irreversibly on the oil drop – otherwise, the polymer would be rinsed away like the surfactant (Fig. 1). In this case, the main role of surfactant is to stabilize the dispersed oil drops and polymer aggregates in the formulation. At low surfactant concentrations, drop flocculation due to polymer bridging takes place [14]. Although most of surfactant is washed away upon rinsing, a part of it remains adsorbed on the oil drops and could influence their attachment to the substrate.

Recently, in experiments with a model shampoo system (based on lauryl sarcosine triethanolamine, CAPB and cationized cellulose) Akiyama et al. [24] established the formation of submicron-sized particles mainly consisting of the cationized polymer in water, which do not adsorb on the hair surface at NaCl concentrations below 0.20 wt %. For 0.75–3.00 wt % NaCl, the polymer-surfactant complex particles were found to easily adsorb on hair. These experiments demonstrate that the electrolyte (NaCl) can also play an important role in the deposition processes with shampoo systems.

The amount of deposited oil has been determined by means of Fourier transform infrared spectroscopy [24]; atomic absorption spectroscopy [25]; fluorescence microscopy [26]; X-ray fluorescent spectroscopy [9,10]; ellipsometry [12], and X-ray analysis system with scanning electron microscope [6]. These methods characterize the total amount of oil as a final result of the deposition process. They do not give information for the occurrence of this process, and especially, for the degree of dilution, at which the oil-drop deposition begins. One method for detailed study of the drop deposition in surfactant + polymer solutions is to immobilize an oil drop to the cantilever of an atomic-force microscope (AFM) and to investigate the drop/substrate interaction force [27].

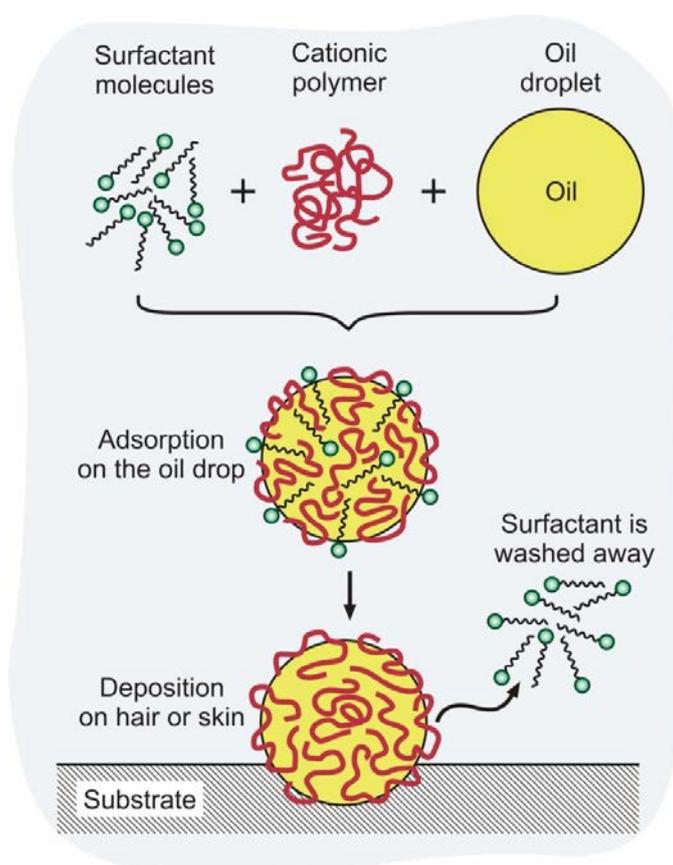

**Fig. 1.** Sketch of the oil-drop deposition process.

In the present study, we propose another method for investigation of the oil-drop attachment to a solid substrate, which gives information about the occurrence and mechanism of this process. The experimental setup is similar to that used for capillary-bridge dynamometry (CBD) [28]. An oil drop is formed at the tip of a capillary in a polymer +

surfactant solution. The drop is first pressed to the substrate and then detached from it. The drop profile during the detachment indicates whether there is (or there is not) drop adhesion to the substrate. The experiment is repeated at different degrees of dilution, which allows one to determine the threshold concentration for drop adhesion, $C_{adh}$. As demonstrated below, by using this method we can determine the effect of the type of the used anionic surfactant; of the composition of the anionic/zwitterionic surfactant mixture, and of other surfactant and salt additives on $C_{adh}$. The results from these experiments are compared with those for the deposition of free oil drops from an emulsion.

It should be noted that the commonly used cationic deposition polymers irreversibly adsorb on solid substrates [30], which are negatively charged, including hair [31] and skin [32]. In this way, the drop deposition happens on a solid surface that has been already modified by the adsorbed polymer. For this reason, it is possible to carry out drop deposition experiments with model substrates, which are representative for hair and skin coated with adsorption layer from the same polymer. This is confirmed by our experiments with hydrophilic and hydrophobized glass substrates, which exhibit similar tendencies with respect to the oil-drop deposition in the presence of pre-adsorbed polymer; see Sections 3.4 and 3.6. In particular, the contact angle (against water) of the hydrophobized glass substrates is in the same range as for hair; see Section 3.1.

## 2. Materials and methods

### 2.1. Materials

The sulfonated methyl esters (SME) of myristic and palmitic acids (C14 and C16) were produced by the Malaysian Palm Oil Board (MPOB) and KLK OLEO; see Ref. [33] for the structural formula of C$n$-SME and its conformation in adsorption layers. C14-SME ($M_w$ = 344 g/mol) and C16-SME ($M_w$ = 372 g/mol) were supplied as dry powders and used in our experiments without additional purification. They are known also as $\alpha$-sulfo fatty methyl ester sulfonates ($\alpha$-MES). The critical micelle concentrations (CMC) of C14- and C16-SME obtained by electric conductivity measurements are, respectively, 4.0 and 1.1 mM [34]. The amounts of water in the SME samples were determined by Karl Fischer analysis and taken into account when calculating the solutions' concentrations.

Another used anionic surfactant was sodium laurylethersulfate (SLES) with one ethylene-oxide group, product of Stepan Co.; commercial name STEOL CS-170; molecular mass 332.4 g/mol. The critical micellization concentration of STEOL CS-170 is 0.7 mM determined by both surface tension and conductivity measurements at 25 °C [35-37]. STEOL

CS-170 contains alkyl chains in the range C10–16, which is the reason for its lower CMC. Note that the CMC of the pure sodium laurylethersulfate is 3 mM [38].

The used zwitterionic surfactant was cocamidopropyl betaine (CAPB), product of Evonik; commercial name Tego® Betain F50; molecular mass 356 g/mol. The CMC of CAPB is 0.09 mM determined by surface tension measurements at 25 °C [35]. As an additive, we used also the nonionic surfactant coco-fatty-acid-monoethanolamide (CMEA), supplied by KLK OLEO, with an average molecular weight $M_w \approx 257.4$ g/mol. The cationic polymer was Jaguar® C-13-S (guar hydroxypropyltrimonium chloride), a high molecular weight polymer product of Solvay, which will be further denoted "Jaguar-C13S" for brevity. The ionic strength was varied by the addition of sodium chloride, NaCl (Sigma, Germany).

The mixed solutions of SLES + CAPB find wide applications in hair-care and skin-care products because of the synergistic growth of wormlike micelles in their mixed solutions, which gives rise to a significant rise of viscosity [39,40,41]. Recently, analogous effects were found in the mixed solutions of C14- and C16-SME with CAPB, which makes the sulfonated methyl esters promising ingredients for personal-care formulations [42]. The sulfonated methyl esters are known also as high performance surfactants in laundry [43] and under hard water conditions [44].

The aqueous solutions were prepared with deionized water purified by Elix 3 water purification system (Millipore). All experiments were carried out at a temperature of 25 °C. The $\zeta$-potential of silica particles (Excelica UF305, Tokuyama Co., Japan) were measured using apparatus Zetasizer Nano ZS (Malvern Instruments, UK).

The used sunflower seed oil (SSO) was a food grade commercial product from a local producer, which was purified by passing through a column filled with adsorbents Florisil and Silica gel 60. Three consecutive passages were applied in order to obtain oil that was free of polar contaminants, as indicated by the constancy of the interfacial tension, which was 30.0 ± 0.5 mN/m, which is close to the value of 32 mN/m obtained by other authors [45]. The viscosity of SSO is 49.1.3 mPa·s and its density is 0.919 g/cm$^3$ (25 ºC). We used also silicon oil, vinyl terminated polydimethylsiloxane of kinematic viscosity 100 cSt and mass density 0.97 g/cm$^3$ (Gelest Inc.).

## 2.2. The pressed drop method (PDM)

To investigate the effect of surfactant/polymer concentration on the deposition (adhesion) of oil drops on solid substrates in aqueous solutions, we constructed the following setup (Fig. 2). The oil drops were formed at the tip of a metal capillary (hollow needle) of outer diameter 1.83 mm. The metal capillary was mounted in a Drop Shape Analyzer, DSA30, apparatus (Krüss GmbH, Germany) upgraded with a piezo-driven membrane for a precise control of the drop volume. Upon increasing its volume, the drop was pressed against the substrate, which is placed on the bottom of a glass cuvette filled with the investigated aqueous solution. Different degrees of dilution of the initial solution were realized by exchange of portions of the solution with pure water. This was realized by two syringes: one of them takes away a given volume of the *solution*; next the other one supplies the same volume of *water*, so that the volume of the aqueous phase in the cuvette remains constant (Fig. 2). After the drop adhesion experiment (see below), the operation with the two syringes is repeated to reach the next degree of dilution, at which the next drop adhesion experiment is carried out. This procedure of *stepwise dilution* is repeated many times. (In the experiments aimed at proving the irreversible adsorption of the cationic polymer on the drop surface, we used another procedure of *complete exchange* of the polymer solution with water, which is described in Appendix A.)

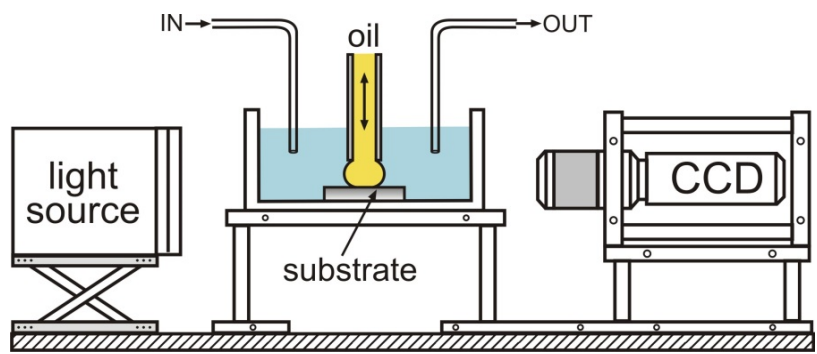

**Fig. 2.** The pressed drop method: Setup for studying oil-drop adhesion to a substrate (details in the text).

The experimental procedure of the *pressed drop method* (PDM) consists of the following steps:

(1) Initially, the cuvette was filled with a concentrated solution of surfactant or surfactant + polymer. In our experiments, the initial total concentration of SLES + CAPB was 6 wt% and the initial polymer concentration was 0.1 wt% Jaguar-C13S. Next, an oil drop was formed at the tip of capillary in the solution, pressed to the substrate and left in contact with the substrate for 10 min. After that, the oil-drop volume was gradually decreased by sucking of oil (at fixed distance between the end of the capillary and the substrate). The variation of the shape of the drop during its detachment indicates whether it has been adherent to the substrate (see below). The drop can be pressed and detached from the substrate several times to check the reproducibility of the experimental result. Repeated experiments with the same drop and with newly formed drops showed that the results are reproducible.

(2) The oil drop was kept in contact with the plate and the solution was diluted by using injection and sucking of water by the syringes. (The pressing of the drop to the substrate was necessary to prevent drop detachment from the capillary by the hydrodynamic flow.) Furthermore, the drop was left in contact with the solid substrate in the diluted solution for 3 min. Next, detachment was tried again to detect drop adhesion to the substrate, and the reproducibility was verified by several pressing/detachment cycles.

Step 2 was repeated several times at different subsequent degrees of dilution.

The time of drop contact with the substrate at step 1 (10 min, deposition) was chosen to be longer than that at step 2 (3 min, rinsing) in order to mimic (to some extent) the real washing process.

In our experiments the substrates were either hydrophilic microscope slides (glass plates), or the same slides subjected to hydrophobization (silanization) by hexamethyldisilazane (HMDS). To characterize the degree of hydrophilicity / hydrophobicity of the plates, a drop of pure water was placed on their surface and their three-phase contact angle was measured goniometrically, by side-view observations using the devise DSA30 (Krüss GmbH, Germany).

As already mentioned, the setup of the pressed drop method (Fig. 2) is similar to that of the capillary bridge dynamometry (CBD) [29]. In CBD, the pressure in the drop is measured by pressure transducer and the drop profile is processed by computer to determine the adhesion (capillary-bridge) force as a function of time. In this respect, the pressed drop method (PDM) is much simpler, because neither pressure measurements nor drop-profile processing is necessary.

In PDM, it is essential that the solution's concentration can be decreased by consecutive steps. At the higher surfactant concentrations, no drop-to-substrate adhesion is observed, whereas at the lower concentrations adhesion can be observed. The results of the PDM measurements are experimental dependencies of the *threshold concentration for drop adhesion* (for brevity called "adhesion concentration") on the solution's composition, e.g. on the SLES/CAPB ratio (see below). The higher the adhesion concentration, the easier the oil-drop deposition on the substrate.

*2.3. Comparative experiments on emulsion drop deposition*

In the PDM measurements, the oil drops have millimeter size, whereas the drop diameter in emulsions is usually of the order of micrometers, and even smaller. To verify whether the results obtained by PDM measurements are representative also for oil-drop deposition from emulsions, we carried out experiments with the same substrates, hydrophilic and hydrophobized glass plates, immersed in the emulsion.

For this goal, oil-in-water emulsions were prepared by using an Ultra-Turrax rotor-stator homogenizer at different concentrations of polymer and surfactant, but at the same volume fraction of oil, viz. 1 %. Because of the relatively low volume fraction of oil, the used input surfactant concentrations were high enough to stabilize these emulsions.

The *stock emulsion* contained 10 vol% silicone oil dispersed in 20 mM aqueous solution of C16-SME. First, a premix was obtained by using the homogenizer at 13,500 rpm for 5 min; next, the final emulsion was obtained at 24,000 rpm for 3.5 min. The *initial solution* contained 6 wt% C16-SME and 0.1 wt% Jaguar-C13S. The experimental procedure consists of the following steps.

(1) Several glass plates (5 in our experiments) were positioned along the cylindrical inner wall of a sufficiently large laboratory glass beaker. The plates were almost vertical, leaning on the wall of the beaker and being completely immersed in the initial solution of volume 40 mL. Next, 4 mL of the stock emulsion was poured in the beaker and gently homogenized by a magnetic stirrer, which resulted in ≈1 vol% oil in the obtained emulsion. After 3 min of plate/emulsion contact, an additional amount (200 mL) of the initial solution was poured in the beaker to remove the formed emulsion cream by overflowing of liquid phase from the beaker. Then, one of the plates was quickly taken out of the liquid. Thus, we avoided additional deposition of oil drops by the receding meniscus during the passage of the plate across the cream [46]. Next, the plate was gently rinsed with pure water to remove weakly attached oil drops and was subjected to microscope observations. (With this step, we mimic the rinsing of hair – only the oil drops, which remain attached after rinsing, matter for hair conditioning.) Because the volume of the initial solution poured during the overflowing (200 mL) was much larger than the volume of the 1 vol% o/w emulsion (44 mL), the rest of the plates in the beaker remained immersed in the initial solution (with a negligible amount of residual oil drops).

(2) The solution in the beaker was diluted to the next desired concentration and the volume of the obtained *new solution* in the beaker was reduced to 40 mL. Next, 4 mL of the stock emulsion was poured in the beaker and gently homogenized by a magnetic stirrer. After 3 min of plate/emulsion contact, an additional amount (200 mL) of the new solution is poured in the beaker to remove the formed emulsion cream by overflowing of liquid phase from the beaker. Then, another plate was quickly taken out of the liquid, rinsed with pure water to remove weakly attached oil drops and subjected to microscope observations.

Step 2 was repeated several times. Thus, a series of plates that have been in contact for a given time with emulsions of the same oil volume fraction (1 vol%), but at different (decreasing) surfactant concentrations can be investigated, and it is possible to determine the threshold concentration for drop adhesion (if any).

It should be noted that the adsorption of the surfactant on the substrate is expected to be reversible, whereas the adsorption of the larger polymer molecules is irreversible [30]. For this reason the plates were kept in contact with the aqueous phase at all preceding steps of dilution to mimic the experimental procedure of PDM; see section 2.2.

## 3. Experimental results and discussion

### 3.1. Contact angle and zeta-potential measurements

The mean contact angles from measurements with many plates are 23.1º ± 5º for the hydrophilic (non-treated) glass plates and 87.7º ± 5º for the hydrophobized glass plates (Fig. 3). For comparison, for hair the advancing contact angles are between 103º (virgin hair) and 70º (chemically damaged hair) [47], i.e. the contact angle of the hydrophobized plate is in the middle of this interval of angles; see also Ref. [48].

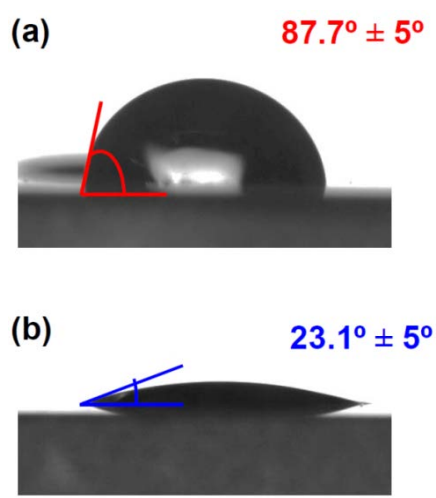

**Fig. 3.** Substrate/water/air contact angles measured from side-view photos of water drops on (a) hydrophobized and (b) hydrophilic glass substrates used in our experiments.

To prove that the used cationic polymer irreversibly adsorbs on oppositely charged solid surfaces, we carried out $\zeta$-potential measurements with silica particles (Excelica UF305, Tokuyama Corp., Japan). In a reference experiment, the particles (as provided by the producer, without additional pretreatment) were dispersed in 10 mM NaCl aqueous solution. The bigger particles sediment at the bottom of the vial. The electrophoretic measurements were carried out with the smaller particles, which remain dispersed, and $\zeta = -20$ mV was measured. In another experiment, the particles were first immersed for 15 min in 0.1 wt% Jaguar-C13S aqueous solution (0.1 wt % is the initial polymer concentration in our dilution experiments). Next, the aqueous phase was removed by centrifugation, and the particles were re-dispersed in 10 mM NaCl aqueous solution (without Jaguar-C13S). In this case, $\zeta = +20$ mV was measured. This rise of $\zeta$-potential (with 40 mV) confirms that Jaguar-C13S strongly and irreversibly adsorbs on oppositely charged solid substrates, as expected [30].

*3.2. Profile of non-adherent and adherent oil drops*

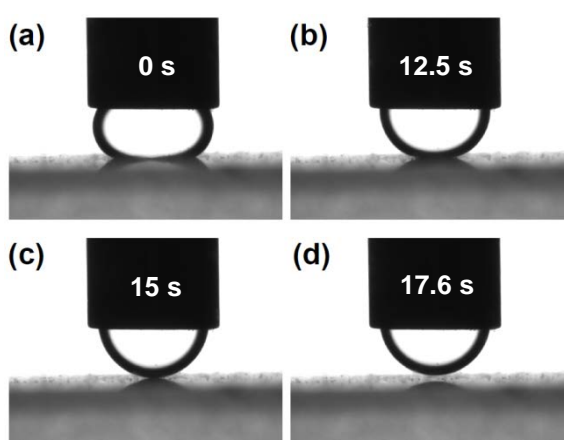

**Fig. 4.** (a-d) Consecutive stages of the detachment of initially pressed oil drop in the *absence* of drop/substrate adhesion. The outer diameter of the capillary is 1.83 mm.

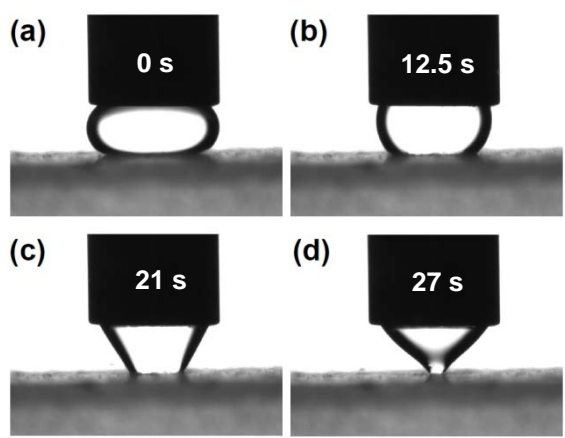

**Fig. 5.** (a-d) Consecutive stages of the detachment of initially pressed oil drop in the *presence* of drop/substrate adhesion. The outer diameter of the capillary is 1.83 mm.

As described in section 2.2, initially the volume of the oil drop is increased and it is pressed against the substrate. This not always leads to drop/substrate adhesion. The presence/absence of adhesion is established by a subsequent decrease of the drop volume (to detach the drop) and observation of the changes in the drop profile, which are different for non-adherent and adherent drops.

In the case of *non-adherent* drop, during the drop detachment the oil/water/solid dynamic contact angle remains very small, close to zero; see Fig. 4 and Supplementary video 1.

In the case of *adherent* drop, during the drop detachment the dynamic contact angle markedly increases and the drop profile becomes more elongated (stretched); see Fig. 5 and Supplementary video 2.

In the PDM measurements, the initial surfactant concentration was relatively high (6 wt% in our case), and it was decreased in a stepwise manner. At each step, detachment of the drop was carried out to check whether the drop was adherent. At the higher surfactant concentrations the drops were non-adherent, whereas at the lower ones – adherent. The first (highest) surfactant concentration, at which indications for drop adhesion (Fig. 5) were observed, was identified with the threshold concentration for drop adhesion, $C_{adh}$. In all PDM experiments, the distance between the end of the capillary and the substrate was kept constant. The drop attachment/detachment was realized only by slowly increasing/decreasing the drop volume by using the piezo-controlled membrane of the setup.

*3.3. Effects of the separate components*

First of all, we investigated the drop adhesion starting with 6 wt% mixed solutions of the surfactants SME and CAPB *without* Jaguar-C13S. In the PDM experiments, we observed that the drops did not adhere to the substrate at all investigated concentrations, except at the lowest one, at which inversion of the contact angle (across water) from acute to obtuse was observed on the hydrophobized plates. The latter indicates direct contact of the oil with the substrate, without surfactant-stabilized aqueous film intervening between them. For example, for an initial concentration of 6 wt% 3:1 C14-SME:CAPB, the adhesion concentration was 0.0078 wt%, at which the initial solution had been diluted 770 times. From the viewpoint of shampoo applications, this is a too high degree of dilution, insofar as most of the oil drops would be driven away during the rinsing with water.

Second, we carried out PDM experiments with solutions of initial concentration 0.1 wt% Jaguar-C13S, *without* surfactants. In this case, the adhesion concentration was 0.004 wt% Jaguar-C13S, at which the initial solution had been diluted 25 times. The drop adhesion could be attributed to adsorbed polymer molecules that are bridging between the oil drop and the substrate. At the higher concentrations, the adsorption layers of the cationic Jaguar-C13S on the negatively charged substrate and oil drop [49] are supposedly thicker and denser, which suppresses the bridging between the oil and substrate by polymer molecules. Instead of bridging, osmotic repulsion between the adsorbed polymer "brushes" could be expected in this case [23]. The positive electric charge of Jaguar-C13S might also contribute to the repulsive force, as indicated by the experiments with added NaCl (see below). As reported in Section 3.1, the adsorption of polymer from 0.1 wt% Jaguar-C13S on the surface of silica particles raises their $\zeta$-potential from $-20$ to $+20$ mV.

The experiments with the oscillating drop method (see Fig. A1 and Table A2 in Appendix 1) indicate that the adsorption of Jaguar-C13S on the oil/water interface is *irreversible* – the surface dilatational modulus remains high, $E' > 10$ mN/m, even if the 0.1 wt% polymer solution is exchanged with pure water. In view of this result, the detection of drop-to-substrate adhesion in the PDM experiment after 25 times dilution could be related to the removal of some weakly attached polymer molecules from the two film surfaces upon dilution.

If 100 mM NaCl is present in the initial solution of 0.1 wt% Jaguar-C13S, then drop adhesion was observed at *all* studied degrees of dilution, including the initial solution. This could be explained with a partial suppression (Debye screening) of the electrostatic attraction between the cationic Jaguar-C13S and the negatively charged drop and substrate. In such a case, the polymer adsorption layers would not be so dense, so that Jaguar-C13S molecules could *bridge* between the drop and the substrate and lead to their adhesion. In addition, electric-field-screening and salting-out effects may lead to enhancement of the short-range hydrophobic *attraction between segments* of polymer molecules adsorbed at the two different surfaces, which also contributes to their adhesion.

These experiments imply that the presence of cationic polymer is a necessary condition for oil-drop deposition in the investigated systems. In other words, Jaguar-C13S plays the role of an adhesive agent.

*3.4. Effect of zwitterionic/anionic surfactant ratio on the oil-drop deposition*

In these experiments, the zwitterionic surfactant was CAPB which was used in mixed solutions with one of the anionic surfactants C14-SME, C16-SME and SLES. The total initial surfactant concentration was fixed to 6 wt%, whereas the weight fraction of CAPB in the surfactant mixture,

$$X_{CAPB} = \frac{w_{CAPB}}{w_{CAPB} + w_{anionic}} \quad (1)$$

was varied. In Eq. (1), $w_{CAPB}$ and $w_{anionic}$ are the weight concentrations of CAPB and the anionic surfactant, respectively. All solutions contained Jaguar-C13S at the same initial concentration of 0.1 wt%. We carried out experiments with both silicon oil and sunflower seed oil (SSO), and with the two types of substrates, hydrophilic and hydrophobized glass (Fig. 3).

As mentioned above, at the higher concentrations the drops were non-adherent (Fig. 4), at the lower concentrations they were adherent (Fig. 5), and the first concentration (upon dilution), at which adherent drop was observed, was identified as the adhesion concentration, $C_{adh}$. In the case of adherent drops, the small value of the initial contact angle (before the onset of drop detachment) indicates the presence of a surfactant-stabilized aqueous film intervening between the oil drop and the substrate.

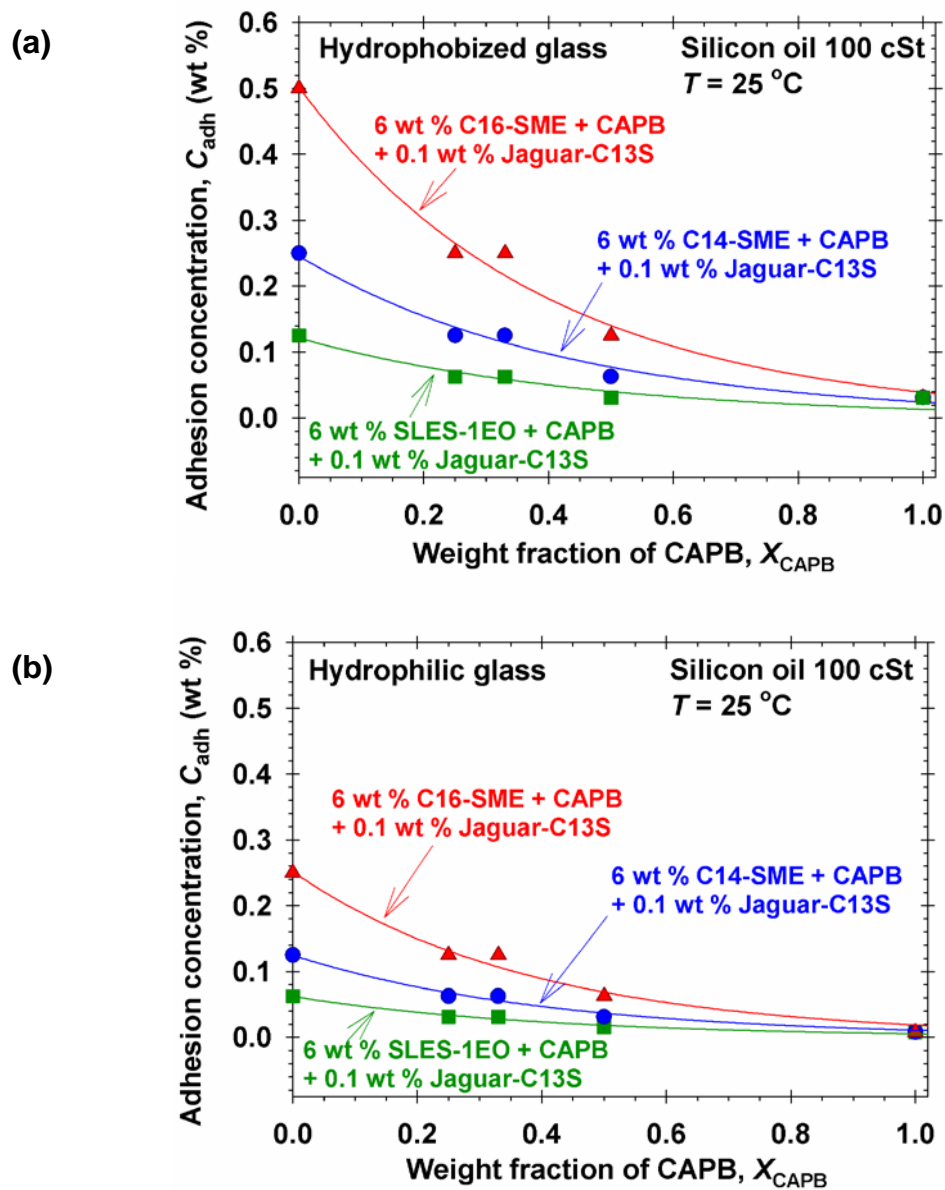

**Fig. 6.** Plots of the adhesion concentration, $C_{adh}$, vs. the molar fraction of CAPB, $X_{CAPB}$, in mixtures with three different anionic surfactants: SLES-1EO, C14-SME and C16-SME. The initial concentrations (before the dilution) have been 6 wt% total surfactant and 0.1 wt% Jaguar-C13S. (a) Hydrophobized and (b) hydrophilic glass substrate. The lines are guides to the eye.

Fig. 6a shows a plot of $C_{adh}$ vs. $X_{CAPB}$ for silicon oil and *hydrophobized*-glass substrate for the three anionic surfactants, C14-SME, C16-SME and SLES. In all cases, the highest $C_{adh}$ (the easiest oil-drop deposition) was observed for 100% anionic surfactant ($X_{CAPB} = 0$), whereas $C_{adh}$ is the lowest at 100% CAPB ($X_{CAPB} = 1$). $C_{adh}$ monotonically decreases with the rise of $X_{CAPB}$. Moreover, $C_{adh}$ is the highest (the oil-drop deposition is the easiest) in the presence of C16-SME and the lowest in the presence of SLES, the values of $C_{adh}$ for C14-SME being intermediate.

For the *hydrophilic*-glass substrate (Fig. 6b), the behavior of the $C_{adh}$-vs.-$X_{CAPB}$ dependences is similar to those for hydrophobized glass (Fig. 6a), only the values of $C_{adh}$ are about 2 times lower. In other words, the oil-drop adhesion to the hydrophilic glass is more difficult in the case of hydrophilic glass ($\theta = 23.1°$, Fig. 3b), which was to be expected.

The interfacial tensions of the two oils against the surfactant + polymer solutions are not identical. For example, for 6 wt% 3:1 C14-SME:CAPB + 0.1 wt % Jaguar-C13S the interfacial tension was 3.50 and 2.25 mN/m for the silicon oil and SSO, respectively. In spite of that, the experiments with drops from silicon oil and SSO gave *identical* results for the $C_{adh}$-vs.-$X_{CAPB}$ experimental curves. This could be explained with the fact that in both cases the oil drop is covered with a dense adsorption layer from surfactant and polymer, so that the substrate interacts with the adsorption layer, rather than with the oil phase itself. In other words, the drop-to-substrate adhesion is governed by attractive forces in the (surfactant + polymer stabilized) aqueous film intervening between the drop and the substrate. Direct contact of the oil with the substrate can take place after the evaporation of water.

*3.5. Discussion*

As already mentioned, the cationic polymer (Jaguar-C13S) adsorbs on both oil/water and solid/water interfaces. Its adhesive role is due (i) to bridging by polymer molecules between the drop and substrate surfaces and (ii) to hydrophobic attraction between hydrocarbon segments of polymer molecules adsorbed at the two different surfaces (Fig. 7). The surfactant binds to the polymer molecules and renders them more hydrophilic, thus, suppressing both the bridging and hydrophobic-segment attraction. For this reason, at the higher surfactant concentrations there is no drop adhesion to the substrate. Upon dilution, the amount of surfactant bound to the polymer decreases, which leads to the appearance of the aforementioned two types of attractive forces and to drop-to-substrate adhesion. At that, upon dilution the anionic surfactant (of higher CMC) is washed out easier than the zwitterionic surfactant. For this reason, the polymer-mediated attraction between the oil drop and substrate is activated earlier at greater fractions of the anionic surfactant, and $C_{adh}$ is the highest at $X_{CAPB} = 0$ (Fig. 6). In contrast, the zwitterionic surfactant (CAPB) of lower CMC binds stronger to the polymer chains and it is more difficult to remove it upon dilution. For this reason, at $X_{CAPB} = 1$ the adhesion concentration, $C_{adh}$, is the lowest, i.e. drop adhesion is observed at the highest degree of dilution.

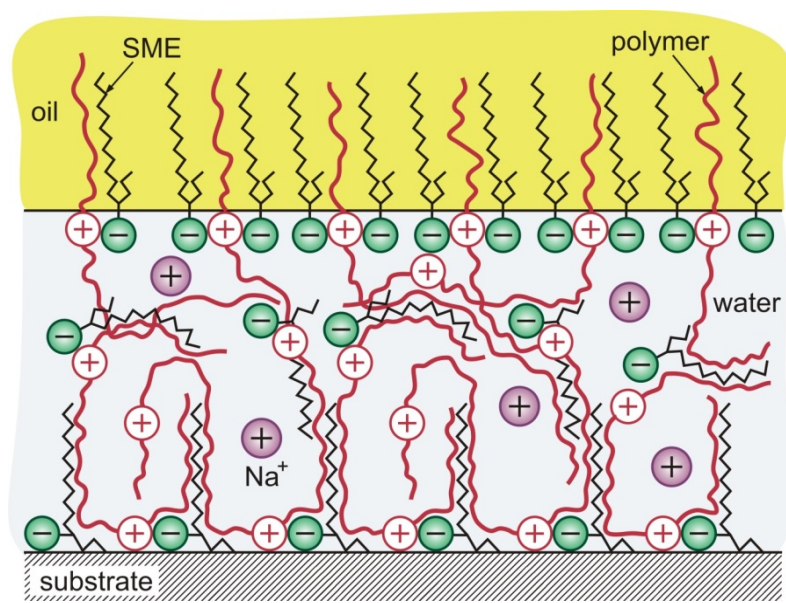

**Fig. 7.** Sketch of a thin aqueous film formed in the contact zone of an oil drop and the substrate in a solution containing a cationic polymer and the anionic surfactant SME. Bridging by polymer molecules between the drop and substrate surfaces and hydrophobic attraction between hydrocarbon segments of polymer molecules adsorbed at the two different surfaces take place. The surfactant binds to the polymer chains and helps for the polymer anchoring to the substrate.

At a first glance, the above explanation contradicts to the fact that the anionic surfactant of the longest hydrocarbon tail (C16-SME) provides the highest $C_{adh}$ (the easiest adhesion upon dilution); see Fig. 6. Indeed, C16-SME is expected to be a stronger hydrophilizer of Jaguar-C13S than C14-SME. However, it seems that the anionic surfactant plays also a second role, different from that of hydrophilizer, viz. the surfactant molecules help for the adsorption of the polymer to the substrate, as sketched in Fig. 7. In this case, the surfactant of longer chain, C16-SME, is expected to have a stronger effect than C14-SME.

The fact that $C_{adh}$ is higher for C$n$-SME than tor SLES could be explained either with a contribution from the methyl group of C$n$-SME to the adhesion energy (Fig. 7), or with different specific binding energies of the sulfonate and sulfate groups to the substrate. For example, it was found that the binding energy of the $Na^+$ counterion to the sulfonate group is higher than to the sulfate group, 2.91 $kT$ vs. 1.64 $kT$ [33].

It should be also noted that around $X_{CAPB} = 0.5$ the solutions become turbid, which is due to the formation of joint aggregates from Jaguar-C13S, CAPB and the anionic surfactant.

These aggregates adsorb on the glass substrate and they are not removed during the supply of water upon dilution. The aggregates neither facilitate, nor impede the oil drop adhesion. Indeed, the $C_{adh}$ vs. $X_{CAPB}$ curves in Fig. 6 have neither maximum nor minimum in the vicinity of $X_{CAPB} = 0.5$.

*3.6. Effects of NaCl and CMEA on the oil drop adhesion*

It is known from the experiment that the addition of NaCl and CMEA to a concentrated mixed micellar solution of anionic surfactant and CAPB leads to increase of its viscosity, i.e. NaCl and CMEA are used as thickeners of shampoo formulations. Here, we investigate what is their effect on the oil-drop deposition. For this goal, using PDM measurements we determined $C_{adh}$ for mixed surfactant solutions containing 0.5 wt% (85.6 mM) NaCl or 0.1 wt% (3.9 mM) CMEA in the *initial* solution (before the dilution); concentrations in the range are used in shampoo formulations. As before, the initial solution contained C16-SME + CAPB at a total concentration of 6 wt% ($X_{CAPB}$ was varied) and 0.1 wt% Jaguar-C13S.

Fig. 8 compares $C_{adh}$-vs.-$X_{CAPB}$ experimental curves for solutions with and without added NaCl and CMEA. Experiments with both hydrophobized (Fig. 8a) and hydrophilic (Fig. 8b) glass substrates were carried out. As before, the experiments with drops from silicon oil and SSO gave identical results for the $C_{adh}$-vs.-$X_{CAPB}$ experimental curves. The data in Fig. 8 exhibit the same tendency as those in Fig. 6. An interesting finding is that the addition of 0.5 wt% NaCl *increases* $C_{adh}$ about twice, whereas the addition of 0.1 wt% CMEA *decreases* $C_{adh}$ almost twice (at $X_{CAPB} = 0$). From the viewpoint of applications, one could expect that the addition of NaCl and CMEA will, respectively, enhance and suppress the oil-drop deposition. The observed effect of NaCl is in agreement with the results in Ref. [24].

The increased adhesion in the presence of NaCl could be explained with the fact that the salt enhances the segment-segment attraction of the hydrocarbon chains in water due (i) to the salting out effect and (ii) to the screening of the electrostatic double-layer repulsion. In contrast, the nonionic surfactant CMEA (like the CAPB) binds to the hydrocarbon chains and renders them more hydrophilic, thus, suppressing the segment-segment hydrophobic attraction and the effect of polymer bridging.

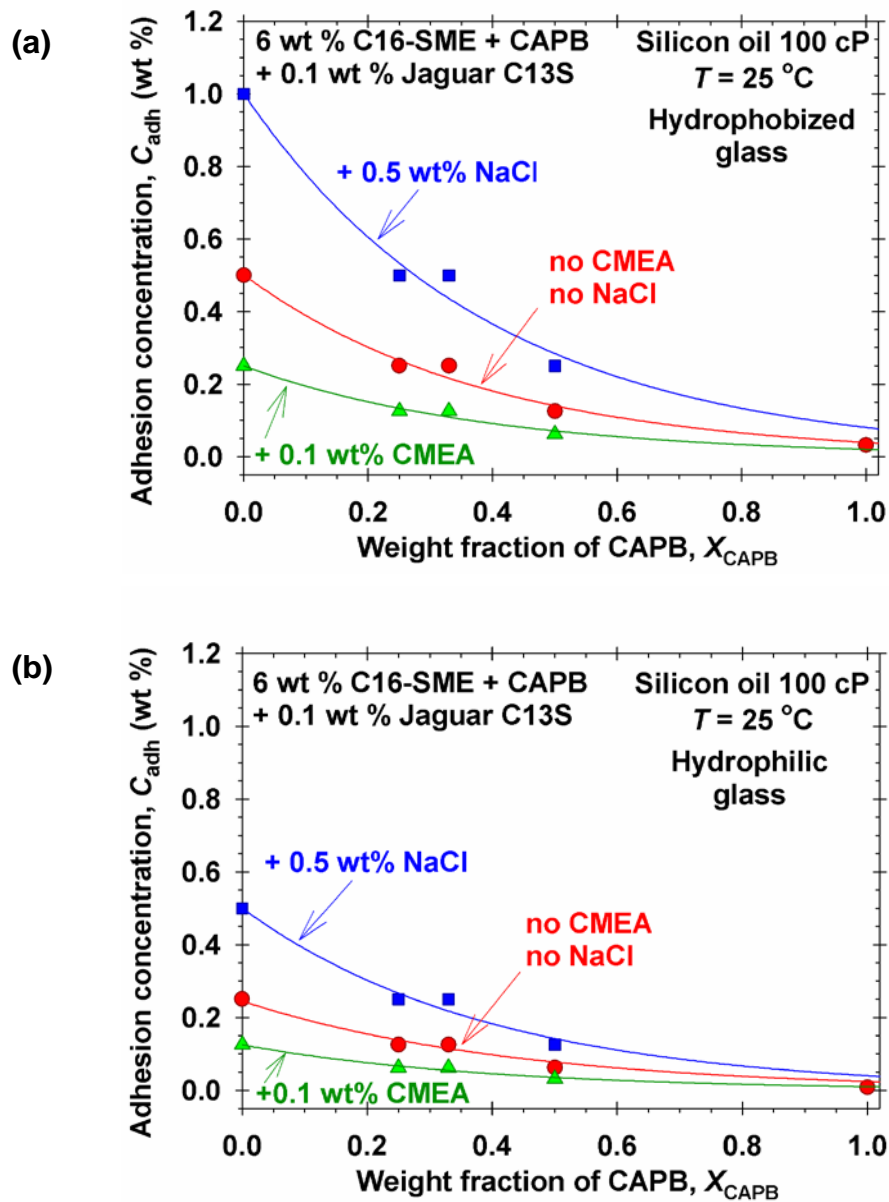

**Fig. 8.** Plot of the adhesion concentration, $C_{adh}$, vs. the molar fraction of CAPB, $X_{CAPB}$, in mixture with C16-SME: effects of added NaCl and coco-fatty-acid-monoethanolamide (CMEA). The initial concentrations (before the dilution) have been 6 wt% C16-SME + CAPB and 0.1 wt% Jaguar-C13S. (a) Hydrophobized and (b) hydrophilic glass substrate. The lines are guides to the eye.

### 3.7. Comparative experiments: drop deposition from emulsion

The results from the comparative experiments with emulsions (see Section 2.3) are shown in Table 1 for hydrophilic substrates and in Appendix A for hydrophobic substrates. These experiments are much more time consuming than the PDM experiments. For this reason, experiments with only one initial solution, 6 wt% C16-SME + 0.1 wt% Jaguar-C13S were carried out. The purpose of these experiments was to verify (i) whether a transition from non-adhering to adhering drops is observed upon dilution, as in the PDM experiments, and (ii) whether the concentration at which this transition happens with emulsions, $C_{adh,em}$, is identical with the adhesion concentration, $C_{adh}$, determined by PDM experiments.

The results for both hydrophilic and hydrophobic substrates (the same as in the PDM experiments, see Fig. 2) show that in the experiments with emulsions a transition from non-adhering to adhering drops is observed upon dilution, just as in the PDM experiments; see Table 1 and Appendix A. In the case of emulsions, the adhesion concentrations for hydrophilic and hydrophobic substrates are $C_{adh,em}$ = 1 and 2 wt% C16 SME, whereas the respective values obtained in the PDM experiments are $C_{adh}$ = 0.25 and 0.5 wt% C16 SME, see Fig. 6. In other words $C_{adh,em} \approx 4C_{adh}$, which means that the drop-to-substrate adhesion happens easier in the experiments with emulsion drops.

A possible explanation of this difference could be the following. The procedure for drop deposition from emulsions was chosen in such a way that the experimental conditions to be as close as possible to those in the PDM measurements. Thus, the glass substrates were hold in the solution during the whole experiment to provide the same adsorption of Jaguar-C13S as in the PDM experiments, having in mind that the adsorption of this polymer could be irreversible. However, differences in the experimental conditions between the two experiments still exist. The experiment shows that equilibrium adsorption from 0.1 wt% Jaguar-C13S solution is reached after 15 min (see Fig. A1 in Appendix A). For this reason, immediately after the addition of the emulsion drops in the investigated solution, they have had a very low surface coverage of Jaguar-C13S (considerably lower than on the equilibrated oil/water interface in the PDM experiments). The interaction of such an emulsion drop with the saturated polymer adsorption layer on the substrate will result in the binding to the drop surface of Jaguar-C13S molecules, which are protruding from the substrate, i.e. in polymer *bridging attraction*. If the adhesion due bridging attraction is stronger and less surfactant-sensitive than the segment-segment attraction (between polymer molecules anchored to the two different surfaces), this could explain the enhanced oil-drop deposition in the experiments with emulsions ($C_{adh,em} > C_{adh}$). The complete explanation of the difference between $C_{adh}$ and $C_{adh,em}$ demands additional investigations, which are out of the scope of the present study.

**Table 1.** Microscope photographs in transmitted and reflected light of hydrophilic glass substrates ($\theta = 23.1°$) that have been immersed in 1 vol% emulsion of silicon oil in water at various concentrations of C16-SME and Jaguar-C13S. The photographs show that the drop deposition (adhesion) begins at $C_{adh,em} = 1$ wt% C16-SME.

| System | Transmitted light | Reflected light |
|---|---|---|
| 6 wt% C16-SME + 0.1 wt% Jaguar | 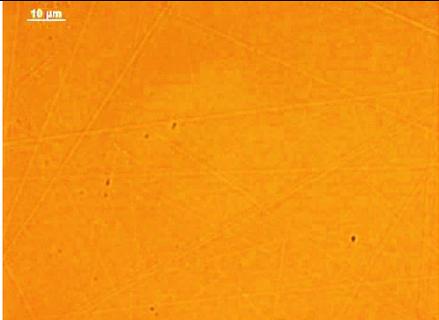 | 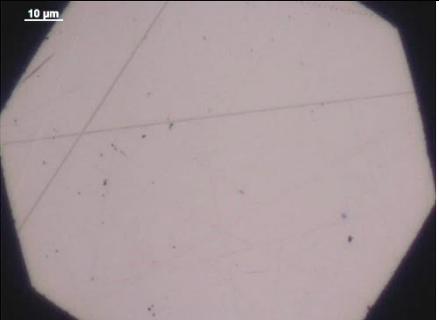 |
| 2 wt% C16-SME + 0.033 wt% Jaguar | 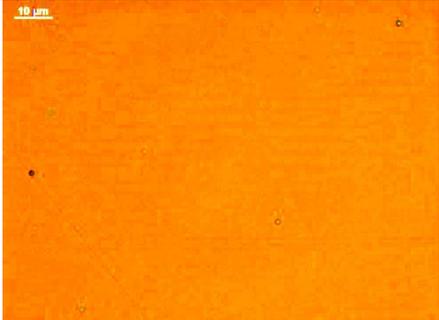 | 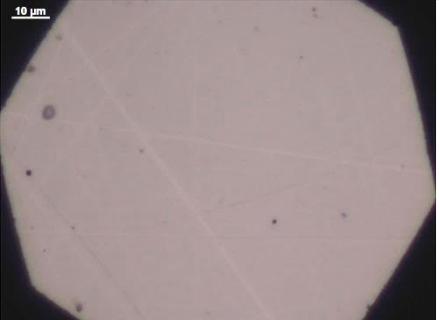 |
| 1 wt% C16-SME + 0.0165 wt% Jaguar | 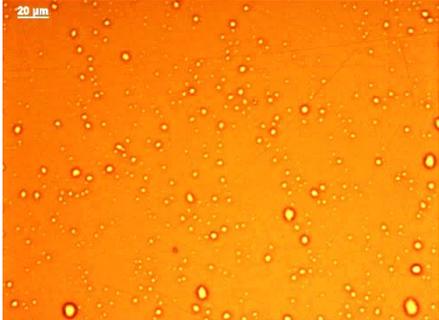 | 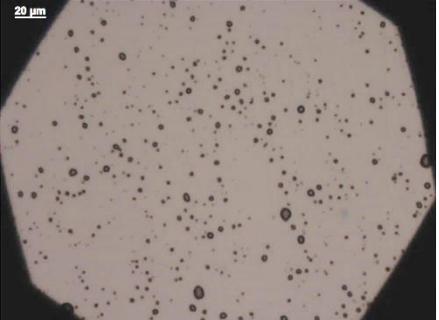 |
| 0.5 wt% C16-SME + 0.008 wt% Jaguar | 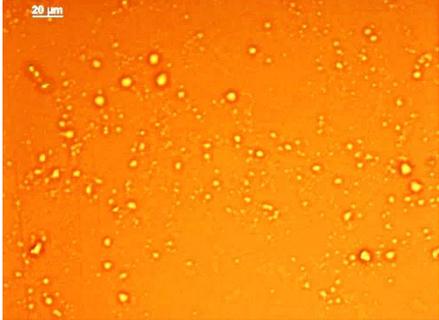 | 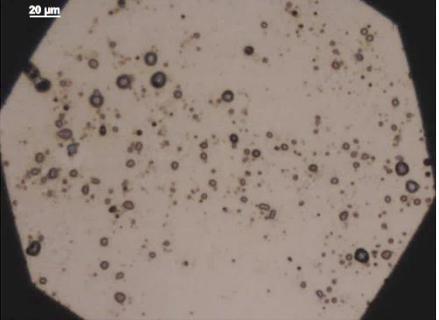 |
| 0.125 wt % C16-SME + 0.002 wt% Jaguar | 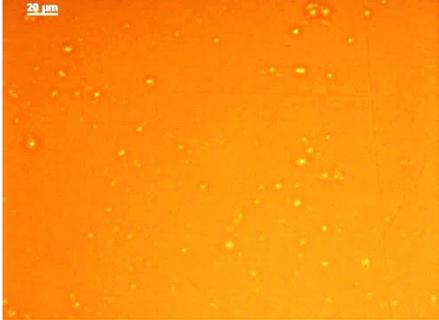 | 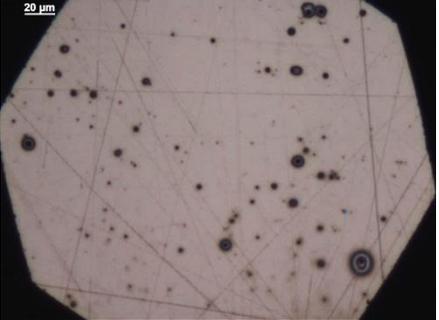 |

## 4. Conclusions

Here, we investigated the deposition of oil drops on solid substrates from mixed solutions of surfactants and cationic polymer, like those used in shampoo formulations. A new method was proposed to study the drop adhesion to substrates of different hydrophobicity. An oil drop is formed at the tip of a capillary; next, pressed against the substrate and, finally, detached from the substrate. The drop profile during the detachment indicates whether there is drop adhesion to the substrate (Figs. 4 and 5). Despite the similarity in the experimental setup, this pressed drop method (PDM) is much simpler than the capillary bridge dynamometry [29], because it is not necessary to measure the capillary pressure and to digitize and process the meniscus profile. The PDM allows one to detect the presence or absence of drop adhesion at different degrees of dilution of the initial solution and, thus, to determine the threshold concentration of drop adhesion, $C_{adh}$. Higher value of $C_{adh}$ corresponds to easier oil-drop deposition. In practical applications, if $C_{adh}$ is lower, a greater portion of the oil drops would be rinsed away before the onset of drop deposition.

In contrast with the other methods [6,9,10,12,25-27], which give the total amount of deposited oil, the PDM experiments give $C_{adh}$ as a function of the solution's composition; see Figs. 6 and 8. Thus, it turns out that the increase of the fraction of CAPB in the mixture with the anionic surfactant suppresses the oil-drop adhesion to the substrate. Moreover, the SME provides easier drop adhesion than SLES. The addition of NaCl enhances the drop deposition, whereas addition of the nonionic surfactant coco-fatty-acid-monoethanolamide, CMEA (often used as shampoo thickener), suppresses the drop deposition. In all cases, the cationic polymer plays a crucial role – no drop adhesion was observed in the absence of polymer. For the adherent oil drops, an aqueous film stabilized by the polymer and surfactant intervenes between the oil and substrate. The drop-to-substrate adhesion is interpreted in terms of the surface forces acting in this film: bridging attraction due to the polymer; hydrophobic attraction between segments of the adsorbed polymer molecules and electrostatic forces; see, e.g., Refs. [14,23]. The results from the PDM experiments are compared with those for the deposition of free oil drops from an emulsion, which confirm the existence of $C_{adh}$ also for drops of smaller size. As a next step, we plan to apply the PDM for optimization of the system with respect to the kind of polymer and its concentration.

The above results evidence that the PDM experiments can be rather informative. They enable one to compare the performance of various components in a personal care formulation and to optimize its composition with respect to the oil-drop deposition.

**Acknowledgements**

The authors gratefully acknowledge the support from KLK OLEO and from the Operational Programme ''Science and Education for Smart Growth", Bulgaria, grant numbers BG05M2OP001-1.002-0023 and BG05M2OP001-2.009-0028.

**Appendix A. Supplementary material**

Supplementary data associated with this article are presented in Appendix A.

# Supplementary Material

for the article

## Oil drop deposition on solid surfaces in mixed polymer-surfactant solutions in relation to hair- and skin-care applications

Rumyana D. Stanimirova [a], Peter A. Kralchevsky [a,*], Krassimir D. Danov [a],

Hui Xu [b], Yee Wei Ung [b], Jordan T. Petkov [b,†]

[a] *Department of Chemical & Pharmaceutical Engineering, Faculty of Chemistry & Pharmacy. Sofia University, 1164 Sofia, Bulgaria.*

[b] *KL-Kepong Oleomas SDN BHD, Menara KLK, Jalan PJU 7/6, Mutiara Damansara, 47810 Petaling Jaya, Selangor Dalur Ehsan, Malaysia*

[†] Present address: *Arch UK Biocides Ltd., Hexagon Tower, Crumpsall Vale, Blackley, Manchester M9 8GQ, UK*

## Appendix A. Additional experimental results

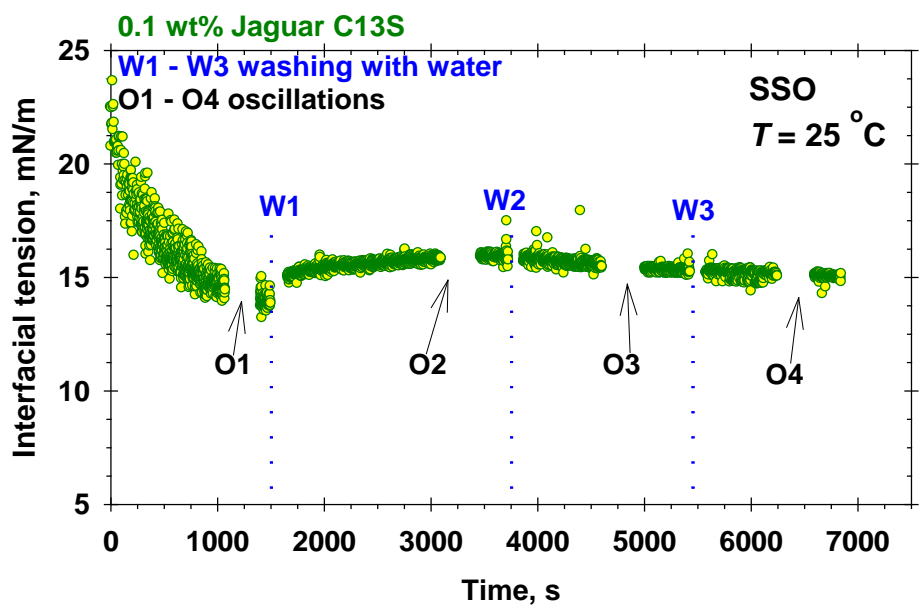

**Fig. A1.** Dependence of the interfacial tension of the boundary between sunflower seed oil (SSO) and aqueous solution of 0.1 wt% Jaguar-C13S vs. time measured with the pendant drop method (with SSO drop). At the moments denoted W1, W2 and W3, the aqueous phase was exchanged with pure water. At the moments denoted O1, O2, O3 and O4, oscillations of the drop volume were applied to determine the interfacial storage and loss moduli, $E'$ and $E''$ (for details, see page 3 below).

**Table A1.** Microscope photographs in transmitted and reflected light of hydrophobized glass substrates ($\theta = 87.7°$) that have been immersed in 1 vol% emulsion of silicon oil in water at various concentrations of C16-SME and Jaguar® C-13-S. The photographs show that the drop deposition (adhesion) begins at $C_{adh,em} = 2$ wt% C16-SME.

| System | Transmitted light | Reflected light |
|---|---|---|
| 6 wt% C16-SME + 0.1 wt% Jaguar | 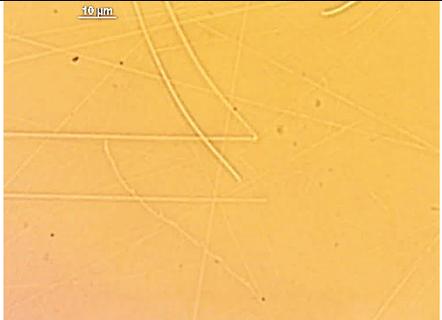 | 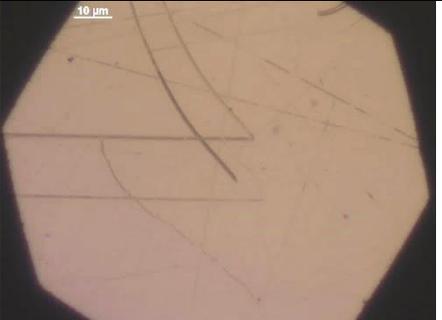 |
| 2 wt% C16-SME + 0.033 wt% Jaguar | 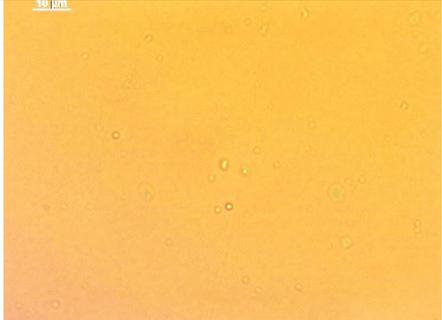 | 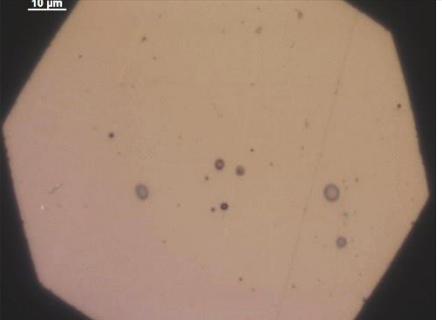 |
| 1 wt% C16-SME + 0.0165 wt% Jaguar | 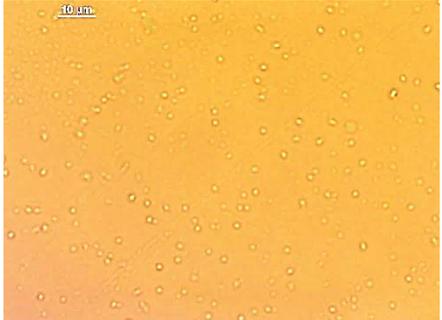 | 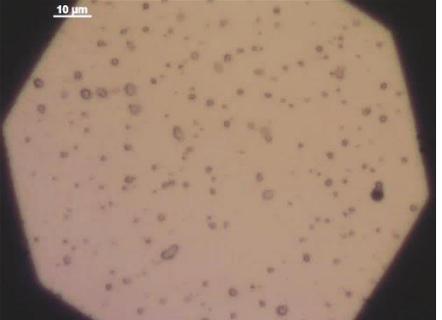 |
| 0.5 wt% C16-SME + 0.008 wt% Jaguar | 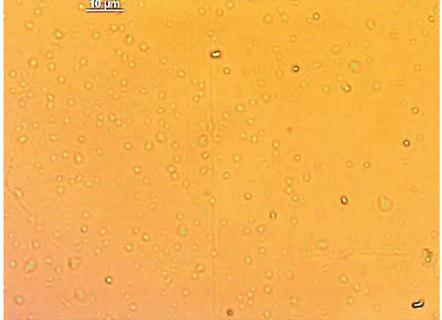 | 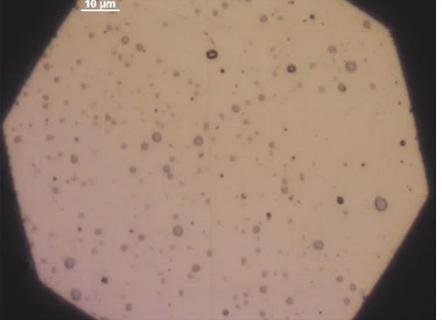 |
| 0.25 wt % C16-SME + 0.004 wt% Jaguar | 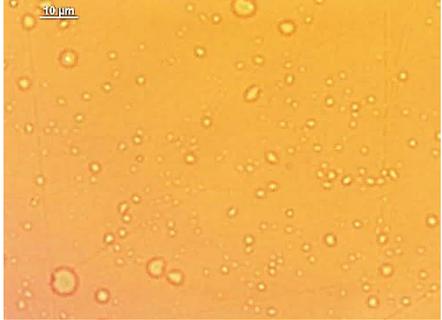 | 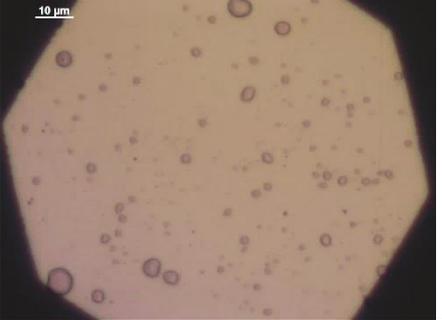 |

The explanations for Table A1 can be found in Sections 2.3 and 3.6 of the main article.

A description of the *oscillating drop method* used to obtain Fig. A1 can be found in the paper S.C. Russev, et al., *Rev. Sci. Instrum.* 79 (2008) 104102.

Fig. 1 shows that the adsorption from a 0.1 wt% Jaguar C13S solution is rather slow: equilibrium adsorption layer is formed after at least 1000 s (more than 15 min). The initial stage corresponds to adsorption kinetics from 0.1 wt% Jaguar C13S solution to the SSO/water interface. The interfacial tension is $\sigma \approx 14.5$ mN/m and the adsorption layer obeys a purely elastic behavior with $E' = 10.5$ mN/m and $E'' \ll E'$ (O1 in Table A2).

Next, we three times exchanged the bulk phases with water (events W1, W2 and W3 in Fig. A1) and carried out oscillations (O2, O3 and O4 in Fig. A1) to measure $E'$ and $E''$. Small increases of the interfacial tension $\sigma$ are due to a small and slow spontaneous (uncontrollable) expansion of the drop-area, $A$. The rheological response of the washed Jaguar C13S layer was purely elastic. This experiment shows that Jaguar C13S *irreversibly adsorbs* at the SSO/water interface.

**Table A2.** Values of the interfacial storage and loss moduli, $E'$ and $E''$, and of the surface area, $A$, of the SSO drop measured in the oscillation experiments O1–O4 in Fig. A1.

| Oscillations | $E'$ (mN/m) | $E''$ (mN/m) | $A$ (mm$^2$) |
|---|---|---|---|
| O1 | 10.5 | 0.1 | 31.2 |
| O2 | 14.8 | 0 | 32.5 |
| O3 | 15.6 | 0 | 32.7 |
| O4 | 16.9 | 0 | 32.7 |

The increase of $E'$ from O1 to O4 could be due (i) to networking in the polymer adsorption layer owing to increasing number of hydrophobic contacts between segments of the polymer chains and/or (ii) to compaction of the polymer adsorption layer promoted by the surface area oscillations. Indeed, during the expansion part of the oscillatory cycle free portions of the oil/water interface might be opened, so that new segments of the adsorbed polymer molecules (initially protruding in the water) may enter the interface.

The exchanges of the bulk phases with water (events W1, W2 and W3 in Fig. A1) were carried out in the following way. The working cuvette had dimensions $35 \times 25 \times 55$ mm. The volume of the solution in the cuvette was 12–13 mL. A cartridge pump (model no. 7523-27, 10–600 RPM, Barnant Co., Cole Parmer Instrument Company, USA) was used to simultaneously supply the new solution and suck out the old one. To exchange the aqueous phase, we ran the pump for 1 min at a flow rate of 150 mL/min. Thus, the volume of liquid in the cuvette was exchanged 12 times for 1 min work of the pump. According to the estimate by Svitova et al. [*J. Colloid Interface Sci*. 261 (2003) 170–179], when the volume of the inserted water reaches 10 times the volume of the cuvette, the solution can be regarded as almost pure water.